\documentstyle[twoside,psfig]{article}
\global\arraycolsep=2pt
\catcode`\@=11
\long\def\@makefntext#1{
\protect\noindent \hbox to 3.2pt {\hskip-.9pt  
$^{{\eightrm\@thefnmark}}$\hfil}#1\hfill}		

\def\thefootnote{\fnsymbol{footnote}}
\def\@makefnmark{\hbox to 0pt{$^{\@thefnmark}$\hss}}	
	
\def\ps@myheadings{\let\@mkboth\@gobbletwo
\def\@oddhead{\hbox{}
\rightmark\hfil\eightrm\thepage}   
\def\@oddfoot{}\def\@evenhead{\eightrm\thepage\hfil
\leftmark\hbox{}}\def\@evenfoot{}
\def\sectionmark##1{}\def\subsectionmark##1{}}

\oddsidemargin=\evensidemargin
\renewcommand{\thefootnote}{\fnsymbol{footnote}}

\newcounter{sectionc}\newcounter{subsectionc}\newcounter{subsubsectionc}
\renewcommand{\section}[1] {\vspace{12pt}\addtocounter{sectionc}{1} 
\setcounter{subsectionc}{0}\setcounter{subsubsectionc}{0}\noindent 
	{\tenbf\thesectionc. #1}\par\vspace{5pt}}
\renewcommand{\subsection}[1] {\vspace{12pt}\addtocounter{subsectionc}{1} 
	\setcounter{subsubsectionc}{0}\noindent 
	{\bf\thesectionc.\thesubsectionc. {\kern1pt \bfit #1}}\par\vspace{5pt}}
\renewcommand{\subsubsection}[1] {\vspace{12pt}\addtocounter{subsubsectionc}{1}
	\noindent{\tenrm\thesectionc.\thesubsectionc.\thesubsubsectionc.
	{\kern1pt \tenit #1}}\par\vspace{5pt}}
\newcommand{\nonumsection}[1] {\vspace{12pt}\noindent{\tenbf #1}
	\par\vspace{5pt}}

\newcounter{appendixc}
\newcounter{subappendixc}[appendixc]
\newcounter{subsubappendixc}[subappendixc]
\renewcommand{\thesubappendixc}{\Alph{appendixc}.\arabic{subappendixc}}
\renewcommand{\thesubsubappendixc}
	{\Alph{appendixc}.\arabic{subappendixc}.\arabic{subsubappendixc}}

\renewcommand{\appendix}[1] {\vspace{12pt}
        \refstepcounter{appendixc}
        \setcounter{figure}{0}
        \setcounter{table}{0}
        \setcounter{lemma}{0}
        \setcounter{theorem}{0}
        \setcounter{corollary}{0}
        \setcounter{definition}{0}
        \setcounter{equation}{0}
        \renewcommand{\thefigure}{\Alph{appendixc}.\arabic{figure}}
        \renewcommand{\thetable}{\Alph{appendixc}.\arabic{table}}
        \renewcommand{\theappendixc}{\Alph{appendixc}}
        \renewcommand{\thelemma}{\Alph{appendixc}.\arabic{lemma}}
        \renewcommand{\thetheorem}{\Alph{appendixc}.\arabic{theorem}}
        \renewcommand{\thedefinition}{\Alph{appendixc}.\arabic{definition}}
        \renewcommand{\thecorollary}{\Alph{appendixc}.\arabic{corollary}}
        \renewcommand{\theequation}{\Alph{appendixc}.\arabic{equation}}
        \noindent{\tenbf Appendix \theappendixc #1}\par\vspace{5pt}}
\newcommand{\subappendix}[1] {\vspace{12pt}
        \refstepcounter{subappendixc}
        \noindent{\bf Appendix \thesubappendixc. {\kern1pt \bfit #1}}
	\par\vspace{5pt}}
\newcommand{\subsubappendix}[1] {\vspace{12pt}
        \refstepcounter{subsubappendixc}
        \noindent{\rm Appendix \thesubsubappendixc. {\kern1pt \tenit #1}}
	\par\vspace{5pt}}

\newcommand{\textlineskip}{\baselineskip=13pt}
\newcommand{\smalllineskip}{\baselineskip=10pt}

\def\eightcirc{
\begin{picture}(0,0)
\put(4.4,1.8){\circle{6.5}}
\end{picture}}
\def\eightcopyright{\eightcirc\kern2.7pt\hbox{\eightrm c}} 

\newcommand{\copyrightheading}[1]
	{\vspace*{-2.5cm}\smalllineskip{\flushleft
	{\footnotesize International Journal of Modern Physics A #1}\\
	{\footnotesize $\eightcopyright$\, World Scientific Publishing
	 Company}\\
	 }}

\newcommand{\publisher}[2]{{\begin{center}\footnotesize\smalllineskip 
	Received #1\\
	Revised #2
	\end{center}
	}}

\def\abstracts#1#2#3{{
	\centering{\begin{minipage}{4.5in}\footnotesize\baselineskip=10pt
	\parindent=0pt #1\par 
	\parindent=15pt #2\par
	\parindent=15pt #3
	\end{minipage}}\par}} 

\renewenvironment{thebibliography}[1]
	{\frenchspacing
	 \ninerm\baselineskip=11pt
	 \begin{list}{\arabic{enumi}.}
	{\usecounter{enumi}\setlength{\parsep}{0pt}
	 \setlength{\leftmargin 12.7pt}{\rightmargin 0pt} 
	 \setlength{\itemsep}{0pt} \settowidth
	{\labelwidth}{#1.}\sloppy}}{\end{list}}

\newcounter{itemlistc}
\newcounter{romanlistc}
\newcounter{alphlistc}
\newcounter{arabiclistc}

\newcommand{\fcaption}[1]{
        \refstepcounter{figure}
        \setbox\@tempboxa = \hbox{\footnotesize Fig.~\thefigure. #1}
        \ifdim \wd\@tempboxa > 5in
           {\begin{center}
        \parbox{5in}{\footnotesize\smalllineskip Fig.~\thefigure. #1}
            \end{center}}
        \else
             {\begin{center}
             {\footnotesize Fig.~\thefigure. #1}
              \end{center}}
        \fi}

\newcommand{\tcaption}[1]{
        \refstepcounter{table}
        \setbox\@tempboxa = \hbox{\footnotesize Table~\thetable. #1}
        \ifdim \wd\@tempboxa > 5in
           {\begin{center}
        \parbox{5in}{\footnotesize\smalllineskip Table~\thetable. #1}
            \end{center}}
        \else
             {\begin{center}
             {\footnotesize Table~\thetable. #1}
              \end{center}}
        \fi}

\def\@citex[#1]#2{\if@filesw\immediate\write\@auxout
	{\string\citation{#2}}\fi
\def\@citea{}\@cite{\@for\@citeb:=#2\do
	{\@citea\def\@citea{,}\@ifundefined
	{b@\@citeb}{{\bf ?}\@warning
	{Citation `\@citeb' on page \thepage \space undefined}}
	{\csname b@\@citeb\endcsname}}}{#1}}

\newif\if@cghi
\def\cite{\@cghitrue\@ifnextchar [{\@tempswatrue
	\@citex}{\@tempswafalse\@citex[]}}
\def\citelow{\@cghifalse\@ifnextchar [{\@tempswatrue
	\@citex}{\@tempswafalse\@citex[]}}
\def\@cite#1#2{{$\null^{#1}$\if@tempswa\typeout
	{IJCGA warning: optional citation argument 
	ignored: `#2'} \fi}}

\def\pmb#1{\setbox0=\hbox{#1}
	\kern-.025em\copy0\kern-\wd0
	\kern.05em\copy0\kern-\wd0
	\kern-.025em\raise.0433em\box0}

\def\fnt#1#2{\footnotetext{\kern-.3em
	{$^{\mbox{\scriptsize #1}}$}{#2}}}

\def\thefootnote{\fnsymbol{footnote}}
\def\@makefnmark{\hbox to 0pt{$^{\@thefnmark}$\hss}}	
	
\def\ps@myheadings{%
    \let\@oddfoot\@empty\let\@evenfoot\@empty
    \def\@evenhead{\slshape\leftmark\hfil}
    \def\@oddhead{\hfil{\slshape\rightmark}}
    \let\@mkboth\@gobbletwo
    \let\sectionmark\@gobble
    \let\subsectionmark\@gobble
    }
\font\tenrm=cmr10
\font\tenit=cmti10 
\font\tenbf=cmbx10
\font\bfit=cmbxti10 at 10pt
\font\ninerm=cmr9

\font\eightrm=cmr8

\def\qed{\hbox{${\vcenter{\vbox{			
   \hrule height 0.4pt\hbox{\vrule width 0.4pt height 6pt
   \kern5pt\vrule width 0.4pt}\hrule height 0.4pt}}}$}}

\renewcommand{\thefootnote}{\fnsymbol{footnote}}  

\newcounter{saveeqn}%
\newcommand{\alpheqn}{\setcounter{saveeqn}{\value{equation}}%
\stepcounter{saveeqn}\setcounter{equation}{0}%
\renewcommand{\theequation}
  {\mbox{\arabic{saveeqn}\alph{equation}}}}%
\newcommand{\reseteqn}{\setcounter{equation}{\value{saveeqn}}%
\renewcommand{\theequation}{\arabic{equation}}}%

\begin{document}
\setlength{\textheight}{7.7truein}  

\thispagestyle{empty}

\markboth{\protect{\footnotesize\it Milton, Kalbfleisch, and 
Gamberg}}{\protect{\footnotesize\it Status of Magnetic Monopoles}}

\normalsize\textlineskip

\setcounter{page}{1}

\copyrightheading{}		

\vspace*{0.88truein}

\centerline{\bf THEORETICAL AND EXPERIMENTAL STATUS}
\vspace*{0.035truein}
\centerline{\bf OF MAGNETIC MONOPOLES}
\vspace*{0.37truein}
\centerline{\footnotesize KIMBALL A. MILTON,\footnote{E-mail address:
{\tt milton@mail.nhn.ou.edu}}\ \,\,  GEORGE R. KALBFLEISCH, and WEI LUO}
\baselineskip=12pt
\centerline{\footnotesize\it Department of Physics and Astronomy, University
of Oklahoma}
\baselineskip=10pt
\centerline{\footnotesize\it Norman, OK 73019-0225,
USA}

\vspace*{10pt}
\centerline{\footnotesize LEONARD GAMBERG}
\baselineskip=12pt
\centerline{\footnotesize\it Department of Physics and Astronomy,
University of Pennsylvania}
\baselineskip=10pt
\centerline{\footnotesize\it Philadelphia, PA 19104-6396, USA}
\vspace*{0.225truein}
\publisher{(received date)}{(revised date)}
\newcommand{\bnabla}{\mbox{\boldmath{$\nabla$}}}
\vspace*{0.21truein}
\abstracts{The Tevatron has inspired new interest in the subject of magnetic monopoles.
First there was the 1998 D0 limit on the virtual production of monopoles,
based on the theory of Ginzberg and collaborators. In 2000 the
first results from an experiment (Fermilab E882) searching 
for real magnetically
charged particles bound to elements from the CDF and D0 detectors were
reported. This also required new developments in theory.  The status
of the experimental limits on monopole masses will be discussed,
as well as the limitation of the theory of magnetic charge at present.
}{}{}


\vspace*{1pt}\textlineskip	
\section{Maxwell's Equations}	
\vspace*{-0.5pt}
\noindent
The most obvious virtue of introducing magnetic charge is the
symmetry thereby imparted to Maxwell's equations,
\alpheqn
\begin{eqnarray}
\bnabla\cdot{\bf E}&=&4\pi\rho_e,\quad \bnabla\cdot{\bf B}=4\pi\rho_m,\\
\bnabla\times{\bf B}&=&{1\over c}{\partial\over\partial t}{\bf E}+
{4\pi\over c}{\bf j}_e,\quad
-\bnabla\times{\bf E}={1\over c}{\partial\over\partial t}{\bf B}+
{4\pi\over c}{\bf j}_m.
\end{eqnarray}
\reseteqn
These equations are invariant under a global {\em duality \/}
transformation.  If $\cal E$ denotes any electric quantity, such as 
$\bf E$, $\rho_e$, or ${\bf j}_e$, while $\cal M$ denotes any magnetic
quantity, such as $\bf B$, $\rho_m$, or ${\bf j}_m$,
the dual Maxwell equations are invariant under
\begin{eqnarray}
{\cal E}\to{\cal E}\cos\theta+{\cal M}\sin\theta,\quad
{\cal M}\to{\cal M}\cos\theta-{\cal E}\sin\theta,
\end{eqnarray}
where $\theta$ is a constant.

J. J. Thomson\cite{jjt}
 (1904) observed the remarkable fact that a {\it static\/}
system of an electric ($e$) and a magnetic ($g$) charge
separated by a distance $\bf R$ possesses an angular momentum,
\begin{eqnarray}
{\bf J}&=&\int (d{\bf r})\,{\bf r\times G}=\int (d{\bf r})\,{\bf r}\times
{{\bf E\times B}\over4\pi c}\nonumber\\
&=&{1\over4\pi c}\int (d{\bf r})\,{\bf r}\times\left[{e{\bf r}\over r^3}
\times{g({\bf r-R})\over({\bf r-R})^3}\right]
={eg\over c}{\bf\hat R},
\end{eqnarray}
which follows from symmetry (the integral can only supply a numerical factor,
which turns out to be $4\pi$).  The quantization of charge 
follows by applying semiclassical quantization of angular
momentum:
\begin{equation}
{\bf J\cdot\hat R}={eg\over c}=n{\hbar\over2},
\quad\mbox{or}\quad eg={n\over2}\hbar c, \quad n=0,\,\pm1,\,\pm2,\,\dots.
\label{thquant}
\end{equation}

\setcounter{footnote}{0}
\renewcommand{\thefootnote}{\alph{footnote}}

\section{Classical Scattering}
\noindent
Actually, earlier in 1896, Poincar\'e\cite{poincare}
 investigated the motion of an electron
in the presence of a magnetic pole.  Let's generalize to two dyons
(a term coined by Schwinger in 1969) 
with charges $e_1$, $g_1$, and $e_2$, $g_2$, respectively.
There are two charge combinations
$q=e_1e_2+g_1g_2$,
$\kappa=-{e_1g_2-e_2g_1\over c}$.
Then the classical equation of relative motion is ($\mu$ is the reduced mass
and $\bf v$ is the relative velocity)
\begin{equation}
\mu{d^2\over dt^2}{\bf r}=q{{\bf r}\over r^3}-\kappa{\bf v}\times{{\bf r}
\over r^3}.
\end{equation}
The constants of the motion are the energy and the angular momentum,
\begin{eqnarray}
E={1\over2}\mu v^2+{q\over r},\quad
{\bf J}={\bf r\times\mu v}+\kappa{\bf \hat r}.
\label{enandang}
\end{eqnarray}
Note that Thomson's angular momentum is prefigured here.

Because $\bf J\cdot \hat r=\kappa$, the motion is confined to a cone,
as shown in Fig.~\ref{fig3}.
\begin{figure}
\centering
\begin{picture}(200,100)
\multiput(0,0)(20,0){10}{\line(1,0){10}}
\put(100,0){\vector(0,1){100}}
\put(85,90){$\bf J$}
\put(100,0){\vector(1,1){50}}
\put(140,25){$\kappa{\bf\hat r}$}
\put(125,80){$\bf \ell=r\times\mu v$}
\put(150,50){\vector(-1,1){50}}
\qbezier[20](100,0)(75,25)(50,50)
\qbezier[30](50,50)(100,70)(150,50)
\qbezier[30](50,50)(100,30)(150,50)
\put(35,15){$\chi/2$}
\qbezier(80,0)(80,7.5)(85,15)
\end{picture}
\vspace*{13pt}
\fcaption{The relative motion of two dyons is confined to the surface of
a cone about the direction of the angular momentum.}
\label{fig3}
\end{figure}
Here the angle of the cone is given by
\begin{eqnarray}
\cot{\chi\over2}&=&{l\over|\kappa|},\quad l=\mu v_0b,
\end{eqnarray}
where $v_0$ is the relative speed at infinity, and $b$ is the impact parameter.
The scattering angle $\theta$ is given by
\alpheqn
\begin{eqnarray}\cos{\theta\over2}&=&\cos{\chi\over2}\left|\sin
\left(\xi/2\over
\cos\chi/2\right)\right|,\quad \mbox{where}\\
{\xi\over2}&=&\left\{\begin{array}{cc}
\tan^{-1}\left({|\kappa|v_0\over q}\cot{\chi\over2}\right),&q>0,\\
\pi-\tan^{-1}\left({|\kappa|v_0\over |q|}\cot{\chi\over2}\right),&q<0.
\end{array}\right.
\end{eqnarray}
\reseteqn
The impact parameter $b(\theta)$ is a multiple-valued function of
$\theta$.
The differential cross section is therefore
\begin{eqnarray}
{d\sigma\over d\Omega}&=&\left|b\,db\over d(\cos\theta)\right|
=\sum_\chi \left(\kappa\over2\mu v_0\right)^2{1\over\sin^4{\chi\over2}}
\left|\sin\chi\,d\chi\over\sin\theta\,d\theta\right|.
\end{eqnarray}
Representative results are given in Ref.~\cite{dyon}.

The cross section becomes infinite in two circumstances; first, when
\begin{equation}
\sin\theta=0 \quad(\sin\chi\ne0), \quad \theta=\pi,
\end{equation}
we have what is called a {\it glory}.
For monopole-electron scattering this occurs for 
\begin{equation}
{\chi_g\over2}=1.047,\,1.318,\,1.403,\,\dots.
\end{equation}
The other case in which the cross section diverges is when
\begin{equation}
{d\theta\over d\chi}=0.
\end{equation}
This is called a {\it rainbow}.
For monopole-electron scattering this occurs at
\begin{equation}
\theta_r=140.1^\circ,\,156.7^\circ,\,163.5^\circ,\,\dots.
\end{equation}
For small scattering angles we have the generalization of the
Rutherford formula
\begin{eqnarray}
{d\sigma\over d\Omega}&=&{1\over(2\mu v_0)^2}\left\{\left(e_1g_2-e_2g_1\over c
\right)^2+\left(e_1e_2+g_1g_2\over v_0\right)^2\right\}
{1\over(\theta/2)^4},\quad\theta\ll1.
\end{eqnarray}

\section{Quantum Theory}
\noindent
Dirac\cite{dirac}
 showed in 1931 that quantum mechanics was consistent with the existence
of magnetic monopoles provided the quantization condition (\ref{thquant}) holds,
which explains the quantization of electric charge.
This was generalized by Schwinger to dyons:
\begin{equation}
e_1g_2-e_2g_1={n\over2}\hbar c.
\end{equation}
(Schwinger sometimes argued that $n$ was an even integer, or even $4$ times
an integer.)

One can see where this comes from by considering quantum mechanical scattering.
To define the Hamiltonian, one must introduce a vector potential, which must
be singular because
${\bf B}\ne\bnabla\times{\bf A}.$
For example, a potential singular along the entire line $\bf \hat n$ is
\begin{eqnarray}
{\bf A(r)}&=&-{g\over r}{1\over2}\left({{\bf \hat n\times r}\over
r-{\bf \hat n\cdot r}}-{{\bf \hat n\times r}\over
r+{\bf \hat n\cdot r}}\right)
=-{g\over r}\cot\theta\,\mbox{\boldmath{$\hat\phi$}}\quad\mbox{if}
\quad{\bf \hat n=\hat z},
\end{eqnarray}
which corresponds to the desired magnetic field from a magnetic monopole,
${\bf B(r)}=g{{\bf r}\over r^3}.$
Invariance of the theory (wavefunctions must be single-valued) under
string rotations implies the charge quantization condition.
 This is a nonperturbative statement.

Yang offered another approach, which is fundamentally equivalent.\cite{yang}
He insisted that there be no singularities, but rather different potentials
in different regions:
\alpheqn
\begin{eqnarray} 
A^a_\phi&=&{g\over r\sin\theta}(1-\cos\theta)={g\over r}\tan{\theta\over2},
\quad\theta<\pi,\\
A^b_\phi&=&-{g\over r\sin\theta}(1+\cos\theta)=-{g\over r}\cot{\theta\over2},
\quad \theta>0.
\end{eqnarray}
\reseteqn
These correspond to the same magnetic field, so must differ by a gradient:
\begin{equation}
A^a_\mu-A^b_\mu={2g\over r\sin\theta}\mbox{\boldmath{$\hat\phi$}}
=\partial_\mu\lambda,
\quad\lambda=2g\phi.\end{equation}
Requiring now that  $e^{ie\lambda}$ be single valued leads to the
quantization condition (\ref{thquant}).

There is also a intrinsic spin formulation, pioneered by Goldhaber.\cite{gold}
The energy (\ref{enandang})
differs by a gauge transformation from
\begin{equation}
{\cal H}={1\over2\mu}\left(p_r^2+{J^2-({\bf J\cdot\hat r})^2\over r^2}\right)
+{e_1e_2+g_1g_2\over r},
\end{equation}
where
\begin{eqnarray}
{\bf J}={\bf r\times p+S}, \quad
\mu{\bf v}={\bf p}+{{\bf S\times r}\over r^2}.
\end{eqnarray}
The quantization condition appears as
${\bf S\cdot\hat r}=m'$.
This was elaborated long ago.\cite{md}

The nonrelativistic Hamiltonian for a system of two interacting dyons is
\begin{equation}
{\cal H}=-{\hbar^2\over2\mu}\left(\nabla^2+{2m'\over r^2}{\cos\theta
\over\sin^2\theta}{1\over i}{\partial\over\partial \phi}-{m^{\prime2}\over
r^2}\cot^2\theta\right)+{q\over r},
\label{nrham}
\end{equation}
where
$m'=-(e_1g_2-e_2g_1)/\hbar c$.
The wavefunction separates:
$\psi({\bf r})=R(r)\Theta(\theta)e^{im\phi}$,
where
\alpheqn
\begin{eqnarray}
\left({d^2\over dr^2}+{2\over r}{d\over dr}+k^2-{2\mu\over\hbar^2}{q\over r}
-{j(j+1)-m^{\prime2}\over r^2}\right)R&=&0,\\
-\left[{1\over\sin\theta}{d\over d\theta}\left(\sin\theta{d\over d\theta}\right)
-{m^2-2mm'\cos\theta+m^{\prime2}\over\sin^2\theta}\right]\Theta&=&j(j+1)\Theta.
\end{eqnarray}
\reseteqn
The solution to the $\theta$ equation is the rotation matrix element:
($x=\cos\theta$)
\begin{eqnarray}
U_{m'm}^{(j)}(\theta)=\langle jm'|e^{iJ_2\theta/\hbar}|jm\rangle
\propto(1-x)^{m'-m\over2}(1+x)^{m'+m\over2}
P_{j-m}^{(m'-m,m'+m)}(x),
\end{eqnarray}
where $P_j^{(m,n)}$ are the Jacobi polynomials, or ``monopole harmonics.''
This forces $m'$ to be an integer.  The radial solutions are
confluent hypergeometric functions,
\alpheqn
\begin{eqnarray}
R_{kj}(r)&=&e^{-ikr}(kr)^LF(L+1-i\eta,2L+2,2ikr),\\
\eta&=&{\mu q\over\hbar^2k},\quad k={\sqrt{2\mu E}\over \hbar},\quad
L+{1\over2}=\sqrt{\left(j+{1\over2}\right)^2-m^{\prime2}}.
\end{eqnarray}
\reseteqn

We solve the Schr\"odinger equation such that a distorted
plane wave is incident,
\begin{equation}
\psi_{\rm in}=\exp i\left[{\bf k\cdot r}+\eta\ln(kr-{\bf k\cdot r})\right].
\end{equation}
Then the outgoing wave has the form
\begin{equation}
\psi_{\rm out}\sim{1\over r}e^{i(kr-\eta\ln2kr)}e^{im'\bar\phi'}f(\theta),
\end{equation}
so up to an unobservable phase, the scattering amplitude is
(here $\theta$ is the scattering angle)
\begin{equation}
2ikf(\theta)=\sum_{j=|m'|}^\infty (2j+1)U^{(j)}_{m'm'}(\pi-\theta)e^{-i(\pi L-
2\delta _L)},
\end{equation}
in terms of the Coulomb phase shift,
$\delta_L=\arg\Gamma(L+1+i\eta)$.
Note that the integer quantization of $m'$ results from the use of an
infinite (``symmetric'') string; an unsymmetric string allows $m'=\mbox{integer}
+{1\over2}$.

One can show that reorienting the string direction gives rise to an
unobservable phase.\cite{md}  Note that this result is completely general:  the
incident wave makes an arbitrary angle with respect to the string direction.
{\em Rotation of the string direction is a gauge transformation.}

Notice that small angle scattering is still given by the Rutherford formula:
\begin{equation}
{d\sigma\over d\Omega}\approx\left(m'\over2k\right)^2{1\over\sin^4\theta/2},
\quad\theta\ll1,\label{quantruth}
\end{equation}
for electron-monopole scattering.  The classical result is good roughly up
to the first classical rainbow.  In general, one must proceed numerically.
Various remarkable results are shown in Ref.~\cite{dyon}.

We can also include the effect of a magnetic 
dipole moment interaction, by adding a term to the Hamiltonian,
\begin{equation}
-{e\hbar\over2\mu c}\gamma\mbox{\boldmath{$\sigma$}}\cdot{\bf H},\quad
{\bf H}=g{{\bf r}\over r^3}.
\label{mmcoupling}
\end{equation}
For small scattering angles, the spin-flip and spin-nonflip cross sections
are (for $\gamma=1$, $\theta\ll1$)
\begin{eqnarray}
\left.{d\sigma\over d\Omega}\right|_{\rm F}&\approx&\left(m'\over2k\right)^2{\sin^2
\theta/2\over\sin^4\theta/2},\quad
\left.{d\sigma\over d\Omega}\right|_{\rm NF}\approx\left(m'\over2k\right)^2{\cos^2
\theta/2\over\sin^4\theta/2}.
\end{eqnarray}
Note that the spin-flip amplitude always vanishes
in the backward direction;
 the spin-nonflip amplitude also vanishes there
for conditions almost pertaining to an electron:
for $m'>0$, and $\gamma=1$, ${d\sigma\over d\Omega}(\pi)=0$.
Various results are shown in Ref.~\cite{dyon}.

\begin{figure}
\vspace*{13pt}
\centerline{\psfig{figure=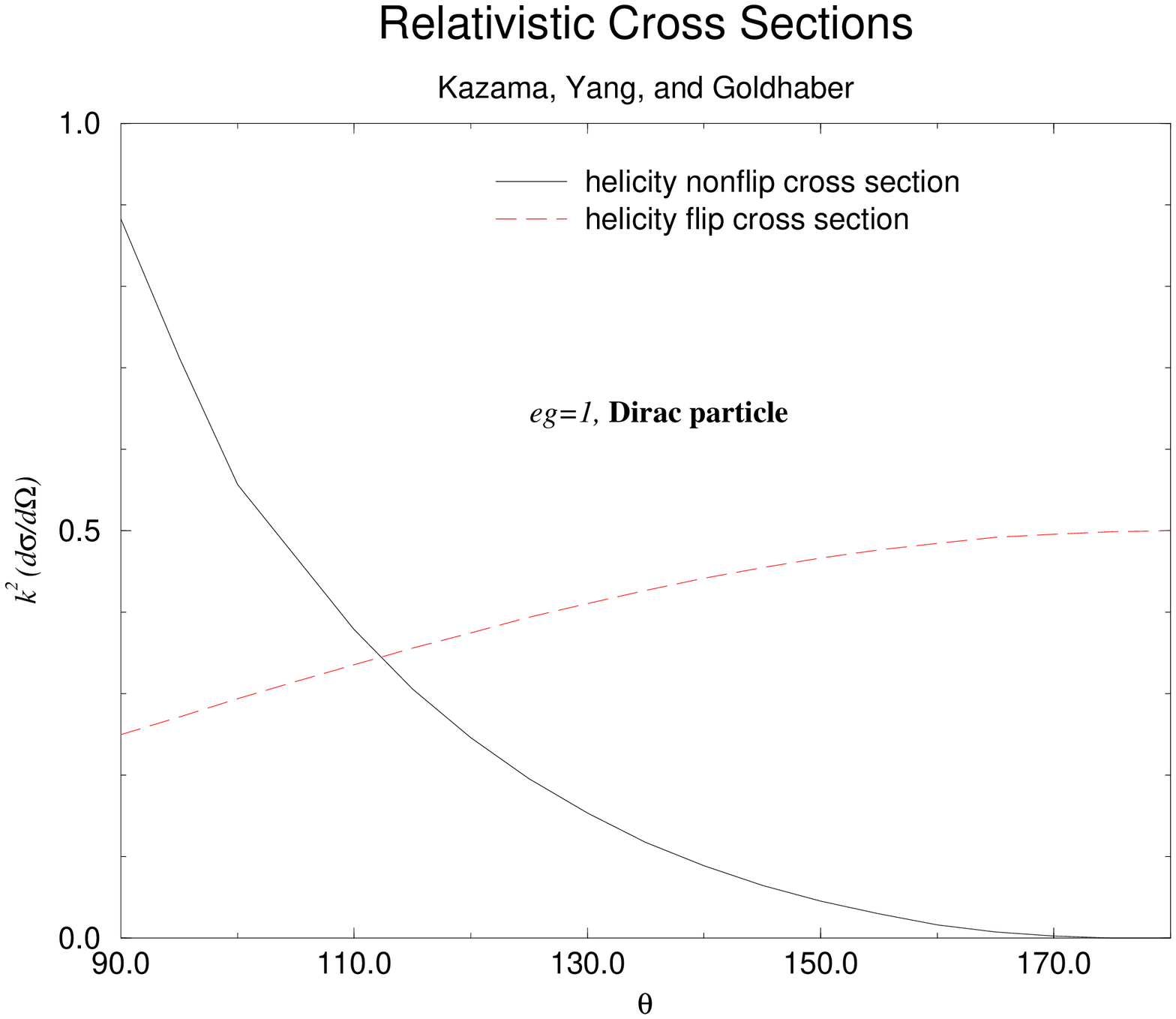,height=3in,width=4in,angle=270}}
\vspace*{13pt}
\fcaption{Relativistic scattering of a spin-1/2 electron by a heavy monopole.}
\label{fig4}
\end{figure}

All of this work was done many years ago in Ref.~\cite{dyon},
which just goes to show that ``good work ages more slowly
than its creators.''  There was of course much earlier work 
\cite{banderet,ford}
A relativistic calculation of the scattering of a spin-1/2 Dirac particle
by a heavy monopole was given by Kazama, Yang, and Goldhaber.\cite{kazama}
  They also gave helicity-flip and helicity-nonflip 
cross sections which are shown in Fig.~\ref{fig4}.
(Note for $\theta=\pi$, helicity nonflip corresponds to spin flip, 
and vice versa.)

\section{Quantum Field Theory}
\noindent
The quantum field theory of magnetic charge has been developed
by many people, notably Schwinger\cite{JS} and Zwanziger.\cite{zwanziger}
A recent formulation suitable for eikonal calculations is given in
Ref.~\cite{gamberg}.
Formal Lorentz invariance is demonstrated provided the quantization
condition holds (rationalized units):
\begin{eqnarray}
\frac{e_{a}g_{b}-e_{b}g_{a}}{4\pi}=
\Bigg\{
\begin{array}{l}
\frac{n}{2}\, , {\mbox{unsymmetric}}\\
n\, , \mbox{symmetric}
\end{array}
\Bigg\}, \quad n \in\, {\mbox{$Z$}}.
\end{eqnarray}
(``Symmetric'' and ``unsymmetric'' refer to the presence or
absence of dual symmetry in the solutions of Maxwell's equations.)

\newcommand{\fg}{\hspace{.1cm}^{\ast}F}
\newcommand{\jg}{\hskip .05cm ^{\ast}\hskip -.07cm j}

The electric and magnetic currents are the sources of the field
strength and its dual:
\begin{eqnarray}
\partial^\nu F_{\mu\nu}=j_\mu \quad {\mbox {\rm and}} \quad
\partial^{\nu}\fg_{\mu\nu}=\jg_{\mu}\ , 
\end{eqnarray}
where 
\begin{equation}
\fg_{\mu\nu}={1\over2}\epsilon_{\mu\nu\sigma\tau}F^{\sigma\tau},
\end{equation}
which imply the dual conservation of electric and magnetic
currents, $j_\mu$ and $\jg_\mu$, respectively,
\begin{eqnarray}
\partial_\mu j^{\mu}=0 \quad {\mbox {\rm and}} \quad
\partial_\mu \jg^{\mu}=0\,  .
\end{eqnarray}
The first-order form of the action describing the interaction of a spin-1/2
electron $\psi$ and a spin-1/2 monopole $\chi$ is
\begin{eqnarray}
W&=&\int (dx) \bigg\{-\frac{1}{2}F^{\mu\nu}(x)
\left(\partial_\mu A_\nu\left(x\right)-\partial_\nu A_\mu\left(x\right)\right)+
\frac{1}{4}F_{\mu\nu}(x)F^{\mu\nu}(x)
\nonumber\\
&+& 
\bar\psi(x)\left(i\gamma\partial+e{\gamma A}(x) 
-m_{\psi}\right)\psi(x)
+\bar\chi(x)
\left(i\gamma \partial+g{\gamma B}(x)-m_{\chi}\right)\chi(x)\bigg\}\, ,
\label{act1}
\end{eqnarray}
where $A_\mu$ and $F_{\mu\nu}$ are independent field variables and
\begin{equation}
B_{\mu}(x)=-\int (dy) f^\nu\left(x-y\right)\fg_{\mu\nu}(y)\, ,
\label{gfb}
\end{equation}
Here $f_\mu(x)$ is 
the Dirac string function which satisfies the differential equation 
\begin{eqnarray}
\partial_\mu f^{\mu}(x)=\delta(x)\, ,
\end{eqnarray}
a formal symmetric solution of which is given by 
\begin{eqnarray}
f^{\mu}(x)=n^\mu\left(n\cdot\partial\right)^{-1}\delta(x)\,,\quad
f^\mu(x)=-f^\mu(-x),
\end{eqnarray}
where $n^\mu$ is an arbitrary vector.  A corresponding dual form in terms
of independent variables $B_\mu$ and ${}^*F_{\mu\nu}$ can be immediately
written down.  Although from these actions a complete path integral
version of dual QED can be given,\cite{gamberg} 
all that is needed for our purposes here is
the relativistic interaction between spinor electric and magnetic currents
$j_\mu=e\bar\psi\gamma_\mu\psi$ and
$\jg_\mu=g\bar\chi\gamma_\mu\chi$:
\begin{equation}
W(j,\jg)=\int (dx) (dx') (dx'')
\jg^\mu(x)\epsilon_{\mu\nu\sigma\tau}
\partial^{\nu} f^{\sigma}\left(x-x^{\prime}\right)
D_{+}\left(x^{\prime}-x^{\prime\prime}\right)j^{\tau}
\left(x^{\prime\prime}\right),
\end{equation}
where the photon propagator is denoted by $D_+(x-x')$.

The modern path integral reformulation of the dual quantum electrodynamics
of electric and magnetic charges were used in Ref.~\cite{gamberg} to
rederive and generalize the eikonal results of Urrutia\cite{urrutia}
for high-energy, low momentum-transfer scattering between an electron
and a monopole.  A simplified version of the argument appears in 
Ref.~\cite{khmono}.
In terms of the quantization condition (\ref{thquant}), the scattering
amplitude, for an arbitrary direction of the incident particle, turns out
to be
\begin{equation}
I({\bf q})=-{4\pi n\over q^2}e^{-2in\phi},
\end{equation}
where $\bf q$ is the momentum transfer.
Squaring this and putting in the kinematical factors we obtain Urrutia's
result\cite{urrutia}
\begin{equation}
{d\sigma\over dt}={(eg)^2\over 4\pi}{1\over t^2}, \quad t=q^2,
\end{equation}
which is exactly the same as the nonrelativistic, small angle result
found in Eq.~(\ref{quantruth}).
This calculation, however, points the way toward a proper relativistic
treatment, and will be extended  elsewhere to the crossed process, the 
production of monopole-antimonopole pairs through quark-antiquark
annihilation.

\section{Previous Searches for Magnetic Monopoles}
\noindent
In the context of ``more unified'' non-Abelian theories, classical composite
monopole solutions were discovered.  The mass of these mono\-poles
would be of the order of the relevant gauge-symmetry breaking scale,
which for grand unified theories is of order
$10^{16}$ GeV
or higher.  But there are models where the electroweak symmetry
breaking can give rise to monopoles of mass $\sim 10$ TeV.
 Even the latter are not yet accessible to accelerator experiments, 
 so limits on heavy monopoles depend either on cosmological 
considerations,\cite{turner}
or detection of cosmologically produced (relic) monopoles
impinging upon the earth or 
moon.\cite{eberhard}
However, {\it a priori}, there is no reason that Dirac/Schwinger monopoles
or dyons of arbitrary mass might not exist: In this respect, it 
is important to set limits below the 1 TeV scale.


 At the University of Oklahoma we are carrying out a
direct search for monopoles produced at the Tevatron, which we will
describe in the next section.
But {\it indirect\/} searches have
been proposed and carried out as well.  De 
R\'ujula\cite{derujula} proposed 
looking at the three-photon decay of the $Z$ boson, where the process proceeds
through a virtual monopole loop. If we use his formula for the
branching ratio for the $Z\to3\gamma$
 process, compared to the current
experimental upper
limit\cite{acciarri}
for the branching ratio of $10^{-5}$, we can rule out
monopole masses lower than about
400 GeV, rather than the 600 GeV quoted by De R\'ujula.
Similarly, Ginzburg and 
Panfil\cite{ginzburg}
and more recently Ginzburg and
Schiller\cite{ginzburg2}
considered the production of two photons with
high transverse momenta by the collision of two photons produced either
from $e^+e^-$ or quark-(anti-)quark collisions. Again the final photons are 
produced through a virtual monopole loop.  Based on this theoretical scheme,
 an experimental limit has 
appeared by the D0 
collaboration,\cite{abbott}
which sets the following bounds
on the monopole mass $M$:
\begin{equation}
{M\over n}>\left\{\begin{array}{cc}
610 \mbox{ GeV}&\mbox{ for } S=0\\
870 \mbox{ GeV}&\mbox{ for } S=1/2\\
1580 \mbox{ GeV}&\mbox{ for } S=1 \end{array}\right.,
\end{equation}
where $S$ is the spin of the monopole.
It is worth noting that a lower
mass limit of 120 GeV  for a Dirac monopole has been 
set by Graf, Sch\"afer, and
Greiner,\cite{graf} based on the monopole
contribution to the vacuum polarization correction to the muon anomalous
magnetic moment. (Actually, we believe that the correct limit, obtained
from the well-known textbook formula for the $g$-factor correction
due to a massive Dirac particle is 60 GeV.)

In Ref.~\cite{khmono} we have criticized these limits on theoretical grounds.
They are based on a naive application of duality, in which the quantization
plays no role.  Thus gauge invariance is not demonstrated.  The Euler-Heisenberg
Lagrangian is used outside its range of validity for hard photon 
processes. That the Euler-Heisenberg Lagrangian is not an effective
Lagrangian in the sense of capturing radiative corrections is demonstrated
by the disparate work of Refs.~\cite{dicus,bordag}.
Moreover, the {\em substitution }
$e\to g$, or
\begin{equation}
\alpha\to\alpha_g={137\over 4}n^2,\quad n=1,2,3,\dots,
\end{equation}
is made, which implies the manifest inconsistency of perturbation
theory (which is already precluded by the nonperturbative quantization
condition). The expansion parameter is $\alpha_g$, which is huge.
 Instead of radiative corrections being of the
order of $\alpha$ for the electron-loop process, these corrections will
be of order $\alpha_g$, which implies an uncontrollable sequence of
corrections.  For example, the internal radiative corrections to the 
four-photon box
diagram have been computed by 
Ritus\cite{ritus} and by
Reuter, Schmidt, and Schubert\cite{reuter} in QED.  In the $O(\alpha^2)$
term in the expansion of the EH Lagrangian,
the coefficients of the $(F^2)^2$ and the 
$(F\,{}^* F)^2$ terms are multiplied by 
$\left(1+{40\over9}{\alpha\over\pi} +O(\alpha^2)\right)$ and 
$\left(1+{1315\over252}{\alpha\over\pi}+O(\alpha^2)
\right)$
respectively.  These corrections become meaningless when we {\it replace\/}
$\alpha\to\alpha_g$.
Moreover, it is easy to see\cite{khmono} that unitarity is violated by
the formulas used in the D0 analysis
unless the monopole masses are above 1 TeV, so the limits quoted are
meaningless.

\section{Oklahoma Experiment: Fermilab E882}
\noindent
The best prior experimental limit on the direct accelerator production
of magnetic monopoles is that
of Bertani et al.\cite{bertani}:
$\sigma\le2\times 10^{-34}\mbox{cm}^2$ for $M\le850\,\mbox{GeV}$.
The fundamental mechanism is supposed to be a Drell-Yan process,
$p+\bar p\to m+\bar m+X$,
where the cross section is given by
\begin{eqnarray}
{d\sigma\over dM}=(68.5n)^2\beta^3{8\pi\alpha^2\over 9s}
\int {dx_1\over x_1}\sum_iQ_i^2q_i(x_1)\bar q_i\left({\cal M}^2\over sx_1\right).
\end{eqnarray}
Here ${\cal M}$ 
is the invariant mass of the monopole-antimonopole pair, and we have
included a factor of  $\beta^3$ 
 to reflect (1) phase space and (2) the velocity
suppression of the magnetic coupling.  
Note that we are unable to calculate the elementary process
$q\bar q\to \gamma^*\to m\bar m$
perturbatively, so we must use nonperturbative estimates.

Any monopole produced at Fermilab loses energy as it passes through the
detector by ionization (computed
using the  energy loss formula of Ahlen\cite{ahlen})
and is trapped in the detector elements with
100\% probability due to interaction with the magnetic moments of the
nuclei.
 The experiment consists of running samples obtained from the old
D0 and CDF detectors through a superconducting induction detector. 
A schematic of our induction detector is shown on the web at
{\tt http://www.nhn.ou.edu/\%7Egrk/apparatus.pdf}.
We are able to set much better limits than Bertani et al.\ because the
integrated luminosity delivered to D0
is $10^4$ larger than that of the previous 1990 experiment:
$\int {\cal L}=172\pm8\,\mbox{pb}^{-1}$.

If $q=e_1e_2+g_1g_2<0$, ${\cal H}_{\rm NR}$ in Eq.~(\ref{nrham})
gives binding of dyons
\begin{equation}
E_{nj}=-{\mu\over2}q^2\left[n+{1\over2}
+\left((j+1/2)^2-m^{\prime2}\right)^{1/2}\right]^{-2}.
\end{equation}
Monopoles will not bind this way---a magnetic moment coupling as in
Eq.~(\ref{mmcoupling}) is required,
in terms the gyromagnetic ratio $\gamma=1+\kappa={g\over2}.$
($\gamma=1$ or $g=2$ is the ``normal'' value.)
The theory\cite{khmono} is somewhat complicated and most
inconclusive, as Table \ref{table} shows.
\begin{table}[ht]
\centering
\tcaption{Weakly bound states of nuclei to a magnetic monopole.
The angular momentum quantum number $J$ of the lowest bound state
is indicated.  In Notes, NR means nonrelativistic and R relativistic
calculations;
hc indicates an additional hard core interaction is assumed, while
FF signifies use of a form factor. IM means induced magnetization, an
additional interaction employed for the relativistic spin-1 calculation.
We use  $n=1$ except for the deuteron, where $n=2$ is required for
binding.}
\begin{tabular}{||c|crcccc||}\hline
\rule[-3mm]{0mm}{8mm}
Nucleus&Spin&$\gamma\,\,\,$&$J$&$E_b$&Notes&Ref\\ \hline
$n$&${1\over2}$&$-1.91$&${1\over2}$&350 keV&NR,hc&\cite{Sivers}\\[0.5ex]
${}_1^1$H&${1\over2}$&2.79&$l-{1\over2}=0$&15.1 keV
&NR,hc&\cite{Bracci}\\[0.5ex]
&&&&320 keV
&NR,hc&\cite{Sivers}\\[0.5ex]
&&&&50--1000 keV&NR,FF&
\cite{Olaussen}\\[0.5ex]
&&&&263 keV&R&
\cite{oooo}\\[0.5ex]
${}_1^2$H&$1$&$0.857$&$l-1=0$ ($n=2$)&${130\over\lambda}$ keV&R,IM
&\cite{Olsen}\\[0.5ex]
${}_2^3$He&${1\over2}$&$-2.13$&$l+{1\over2}={3\over2}$&
13.4 keV&NR,hc&\cite{Bracci}\\[0.5ex]
${}_{13}^{27}$Al&${5\over2}$&3.63&$l-{5\over2}=4$
&2.6 MeV&NR,FF&\cite{Olaussen}\\[0.5ex]
${}_{13}^{27}$Al&${5\over2}$&3.63&$l-{5\over2}=4$
&560 keV&NR,hc&\cite{Goebbel}\\[0.5ex]
${}_{48}^{113}$Cd&${1\over2}$&$-0.62$&$l+{1\over2}={49\over2}$&
6.3 keV&NR,hc&\cite{Bracci}\\[0.5ex]
\hline
\end{tabular}
\label{table}
\end{table}

Unfortunately, the simple theory says that Be (of which the beam pipe is
made)
will not bind to monopoles ($S=3/2$, $\gamma<0$), but the strength
of the magnetic field in the vicinity of a monopole will mostly likely
disrupt the nucleus, ensuring binding.  Al and Pb\footnote{22\%
of naturally occurring Pb is ${}_{\,\,82}^{207}$Pb, which has spin $1/2$
and $\gamma=0.582$, which is sufficient for binding in the lowest
angular momentum state.\cite{khmono}  The other three stable isotopes
have spin 0.}\,\, are okay.

Estimates of binding energies are in the keV range and up.  Simple estimations
show that an energy of an eV is sufficient to bind
a monopole to a nucleus with a
10 year lifetime, and then the monopole-atom complex will remain permanently
bound to the material lattice.  (Evidently, it would not do to melt the
material down.)

A characteristic
 signal would be produced in a superconducting loop contained within
a superconducting can by a magnetic monopole of strength $g$ pulled through it.
Note that if the shield were not present, the supercurrent induced in
the loop of inductance $L$ and radius $R$ would be given by
\begin{equation}
I(t)={2\pi g\over Lc}\left(1-{z(t)\over\sqrt{R^2+z(t)^2}}\right),
\end{equation}
where $z(t)$ is the vertical position of the monopole relative to
the position of the center of the loop.  The theory including the
shield correction can be verified with a  pseudopole.

Background effects are enormous.  All nonmagnetic
but conducting samples possess:
\begin{itemize}
\item Permanent magnetic dipole moments (in the absence of boundaries): 
\begin{equation}
I(t)=-{2\pi\mu_z\over Lc}{R^2\over[R^2+z(t)^2]^{3/2}}
\end{equation}
\item Induced magnetization: ($a$ is the radius of the superconducting
cylinder)
\begin{equation}
I(t)={v\over c^3}{1\over L}\int(d{\bf r})r^2\sigma(r){\partial
B_z\over\partial z'}(z'){1\over R}H\left({z'\over R},{a\over R}\right),
\end{equation}
where $v$ is the velocity with which the sample is pulled through the
detector, $\sigma$ is the conductivity of the sample, and
\alpheqn
\begin{eqnarray}
H\left({z'\over R},{a\over R}\right)&=&\int_0^\infty dy\,y \cos y{z'\over R}
\left[K_1(y)-I_1(y){K_1(ya/R)\over I_1(ya/R)}\right]\\
&\to&{\pi\over 2}{R^3\over(R^2+z^{\prime2})^{3/2}},\quad a/R\to\infty.
\end{eqnarray}
\reseteqn
\end{itemize}

Calibration and real data are shown in the Figures. 
The pseudopole data (Fig.~\ref{fig:fig1})
 clearly shows that we could detect a Dirac pole.
 \begin{figure}[ht]
\centerline{\psfig{figure=Layout0_Fig1AB.EPSF,width=3.4in}}
\fcaption{``Pseudopole'' curves.  
a) Comparison of theoretical monopole response  to an 
experimental calibration and of
 a simple point dipole of one sample with that calculated 
from the theoretical response curve.  b) The observed 
``step'' for a pseudopole current, corresponding to 2.3 minimum 
Dirac poles, embedded in an Al sample.}
\label{fig:fig1}
\end{figure}
As one sees from Fig.~\ref{fig:fig2} real samples have ``large'' 
dipole signals; what we are looking
for is an asymptotic ``step'' 
 indicating the presence of a magnetic charge.
 \begin{figure}[ht]
\centerline{\psfig{figure=Layout0_Fig2ab.EPSF,width=3.4in}}
\fcaption{Sample spectra. a)  
Beryllium sample ``SBe5P,'' and b) aluminum 
sample ``S133Al.'' The observed steps are $-0.8$ mV in a) and 
$+0.4$ mV in b).  The dipole signals are off scale in the middle
regions of the plot in this vertically expanded view.}
\label{fig:fig2}
\end{figure}
Steps seen are typically much smaller than that expected of a magnetic
pole of Dirac strength.  The histogram of steps is shown in
Fig.~\ref{fig:fig3}.
\begin{figure}[ht]
\centerline{\psfig{figure=Layout0_Fig3.EPSF,width=3.4in}}
\fcaption{Histogram of steps.  
Vertical lines (dashed) define the expected positions of signals for 
various $n$.  The Gaussian curve 
(dashed) corresponds to 228 measurements having an average 
value of 0.16 mV and an rms sigma of 0.73~mV.}
\label{fig:fig3}
\end{figure}
For $n=1$ the 90\% confidence upper limit is 4.2 signal events 
for 8 events observed when 10 were 
expected.\cite{feldman}
These
8 samples were remeasured and all fell within $\pm1.47$ mV of $n=0$
(more than 1.28$\sigma$ from $n=\pm1$).
For $n=2$ the 90\% confidence upper limit is 2.4 signal events
for zero events observed and zero expected.
Then, by putting in angular and mass acceptances we can get cross section
limits as shown in Table \ref{tab2}.
\begin{table}
\centering
\tcaption{Acceptances, upper cross section limits, and lower mass limits,
as determined in this work (at 90\% CL).}
\begin{tabular}{ccccc}
Magnetic Charge&$|n|=1$&$|n|=2$&$|n|=3$&$|n|=6$\\
\hline
Sample&	Al&Al&Be&Be\\
$\Delta\Omega/4\pi$ acceptance&0.12&0.12&0.95&0.95\\
Mass Acceptance&0.22&0.060&0.0018&0.11\\
Number of Poles&$<4.2$&$<2.4$&$<2.4$&$<2.4$\\
Cross section limit&0.84 pb&1.9 pb&8.4 pb&0.17 pb\\
Monopole Mass Limit&$>263$ GeV&$>282$ GeV&$>284$ GeV&$>413$ GeV
\end{tabular}
\label{tab2}
\end{table}
These numbers reflect a new analysis, and so differ somewhat from our
published results.\cite{exp} 
To obtain the mass limits, we use the model cross sections referred to
above.


\nonumsection{Acknowledgements}
\noindent
This work was supported in part by the US Department of Energy.
We gratefully acknowledge the help of our
colleagues, particularly Eric Smith and Mike Strauss.
KAM thanks Michael Bordag for asking him to present this talk at QFEXT01.

\end{document}